\documentclass[12pt,preprint]{aastex}

\slugcomment{Ap. J.}

\shorttitle{}
\shortauthors{Chuss et al.}

\begin{document}

\title{Mapping Magnetic Fields in the Cold Dust at the Galactic Center}

\author{David T. Chuss}
\affil{NASA Goddard Space Flight Center}

\author{Giles Novak}
\affil{Department of Physics and Astronomy, Northwestern University}

\author{Roger H. Hildebrand}
\affil{Enrico Fermi Institute, University of Chicago}

\author{C. Darren Dowell}
\affil{California Institute of Technology}

\author{John E. Vaillancourt}
\affil{University of Wisconsin, Madison}

\and 
\author{Jacqueline A. Davidson, Jessie L. Dotson}
\affil{NASA Ames Research Center, Moffett Field, CA}

\begin{abstract} We report the detection of polarized emission in the vicinity of the
Galactic center for 158 positions within eight different pointings of the Hertz
polarimeter operating on the Caltech Submillimeter Observatory. These pointings
include positions 2\arcmin\, offset to the E, NE, and NW of M-0.02-0.07; positions to
the SE and NW of the 20 km s$^{-1}$ cloud (M-0.13-0.08), CO+0.02-0.02, M+0.07-0.08,
and M+0.11-0.08. We use these data in conjunction with previous far-infrared and
submillimeter polarization results to find that the direction of the inferred magnetic
field is related to the density of the molecular material in the following way: in
denser regions, the field is generally parallel to the Galactic plane, whereas in
regions with lower densities, the field is generally perpendicular to the plane. This
finding is consistent with a model in which an initially poloidal field has been
sheared into a toroidal configuration in regions that are dense enough such that the
gravitational energy density is greater than the energy density of the magnetic
field. Using this model, we estimate the characteristic strength of the magnetic
field in the central 30 pc of our Galaxy to be a few mG.

\end{abstract}


\section{Introduction}

Despite significant advancement over the last two decades, the Galactic center
magnetosphere is still not well understood; however, determining the geometry and
strength of the magnetic field in this region is crucial to developing an
understanding of the dynamics of the Galactic center.  One reason that the
magnetosphere is of interest lies in the Galactic center's role as an Active Galactic
Nucleus. In the central engines of galaxies, magnetic fields are believed to be
important in angular momentum transport and jet dynamics.
 
The most striking evidence for the existence of magnetic fields in the Galactic
center is the existence of the Non-thermal Filaments (NTFs) of the Galactic center
Radio Arc (GCRA) \citep{Zadeh84}.  Radio polarization measurements have confirmed the
synchrotron nature of the emission from these filaments and support the notion that
the filaments trace magnetic field lines \citep{Tsuboi86}.  Almost all of the
confirmed NTFs that have been discovered in the Galactic center region are aligned
with their long axes within 20$^\circ$ of perpendicular to the plane of the Galaxy
\citep{LaRosa00}.  The presence of these filaments has led to the idea that these
filaments trace the inner part of a dipole or {\it poloidal} magnetic field.

In many cases NTFs are observed to be interacting with Galactic center molecular
clouds. The best example of this is in the center of the GCRA where the 25 km
s$^{-1}$ molecular cloud associated with G0.18-0.04 (the Sickle) is coincident with
the filaments \citep{Serabyn91}.  The lack of observed distortion of the filaments
allows one to set a lower limit for the strength of the magnetic field. This argument
yields $B>{\mathrm {few\,mG}}$ \citep{Zadeh87}. The association of molecular clouds
with NTFs gives additional evidence for the dynamical importance of these fields and
has motivated the idea that the generation of the relativistic electrons required to
create these NTFs could occur as a result of a molecular cloud/magnetic flux tube
interaction. The specific mechanism thought to be responsible for the generation of
these relativistic electrons is magnetic reconnection between the fields in the flux
tube and those in the cloud (\cite{Serabyn94}; for a review, see \cite{Davidson96}).

Magnetically aligned dust grains are known to emit polarized radiation in the
far-infrared and submillimeter. Thus, polarimetry at these wavelengths has proven to
be a reliable technique for measuring the projected, line-of-sight integrated
magnetic field configuration. Previous polarimetry data have shown that the
field in the dense molecular clouds is not generally consistent with the poloidal
field traced by the filaments. \cite{Novak00} have shown that in the molecular cloud
M-0.13-0.08, the field is nearly parallel to the Galactic plane, indicating that
gravitational rotation and infall have sheared out the field into a {\it toroidal}
configuration. In addition, 60 $\mu$m
polarimetry of the molecular cloud associated with the Sickle indicates that the field in
this region
is parallel to the Galactic plane \citep{Dotson00}.

\cite{Uchida85} have constructed a model that connects poloidal and toroidal fields.
Because the magnetic flux is frozen into the matter, differential rotation and infall
can shear an initially poloidal field into a toroidal one in sufficiently dense
regions of the Galactic center. \cite{Novak02} have shown that on large scales, the
dust in the Central Molecular Zone (CMZ) of the Galaxy is permeated by a field that
is oriented in a direction parallel to the plane of the Galaxy. They compare their
observation to the spatial distribution of the direction of Faraday rotation in the Galactic center
and show that these two different techniques for sampling fields yield results that
are consistent with such a model.

We extend the ideas that underlie the above model to the inner 30 parsecs of the
Galactic center. When resolved with sub-arcminute resolution in the submillmeter
\citep{PiercePrice00}, this region exhibits a clumpiness resulting from spatial
density variations. We find that in less dense regions, the magnetic field that
permeates the dust is poloidal, consistent with that found in the filaments. In
these regions, the matter density is too low for gravity to shear the field into a
toroidal configuration. In the dense molecular clouds, however, the field appears to
be generally toroidal, indicating that in such regions, the energy density of gravity
dominates that of the permeating magnetic field.

Section 2 describes the observations. In section 3, the data are discussed.  
Finally, conclusions are presented in section 4.

\section{Observations}

The data were obtained using the University of Chicago 32-pixel polarimeter, Hertz
\citep{Dowell98} at the Caltech Submillimeter Observatory on Mauna Kea, Hawai'i in
April of 2001. The beam size of Hertz is 20\arcsec\,FWHM with a pixel spacing of
18\arcsec. The central wavelength of the Hertz passband is 350 $\mu$m with a
$\Delta\lambda\over\lambda$ of 0.1.

The observing technique used with Hertz involves a combination of high frequency
``chopping'' in which the array footprint is alternately pointed at the source and a
reference position 6\arcmin\, away in cross-elevation and low frequency ``nodding''
during which the source is alternately placed in the right and left beams of the
telescope. One undesirable complication that this observational scheme introduces is
the problem of polarized flux in the reference beam. This problem is discussed in
detail in \S~\ref{sec:refbeam}.

For a full description of the observing and analysis procedures see
\cite{Schleuning97} and \cite{Dowell98}.

During the observations, a total of 158 new polarization measurements were obtained
having a polarimetric signal-to-noise greater than 3. These data are given in Tables
1 through 7.

\section{Discussion}

\subsection{Morphology of the Inner 30 Parsecs}

The magnetic field vectors inferred from these data are shown in blue in region III
of Figure~\ref{fig:scuba}.  The black vectors in this region are those from
\cite{Novak00}.  Also shown in this figure are 100 $\mu$m data from the Kuiper
Airborne Observatory \citep{Dotson00} of the Arched Filaments (region II) and 60
$\mu$m data \citep{Dotson00} of the Sickle (region I). The 100 $\mu$m and 50 $\mu$m
data have beam sizes of 35$\arcsec$ and 22$\arcsec$, respectively.
The contours trace 850 $\mu$m
flux \citep{PiercePrice00}. Major molecular features are labeled.

\subsubsection{M-0.13-0.08}

The molecular cloud M-0.13-0.08 (commonly called the 20 km s$^{-1}$ cloud) has an
elongated shape, and its long axis is oriented at a shallow (15$^\circ$) angle to the
Galactic plane (See Fig.~\ref{fig:scuba}). The cloud is thought to be located in
front of the Galactic center since it appears in absorption in the mid-infrared
\citep{Price01}. In projection, the long axis of this cloud points in the direction
of Sgr A*.  This fact suggests that this cloud is undergoing a gravitational shear as
it falls toward Sgr A*. This is supported by other observations including a high
velocity gradient along the long axis \citep{Zylka90}, increasing line widths and
temperatures along the cloud in the direction of Sgr A* \citep{Okumura91}, and an
extension of the cloud detected in NH$_3$ emission to within 30$^{\prime\prime}$ of
Sgr A* in projection \citep{Ho91}.  \cite{Novak00} have discovered that the magnetic
field structure is such that the field is parallel to the long axis of the cloud.
They point out that for gravitationally sheared clouds, a consequence of
flux-freezing is that regardless of the initial configuration of the field, the field
will be forced into a configuration in which it is parallel to the long axis of the
cloud. In addition, these authors note that this is not only true for the dense
material inside of the cloud, but also for the more diffuse molecular material that
belongs to the ambient region. They note that the southern end of this cloud exhibits
a flare both morphologically and in magnetic field structure. They suggest that this
flare is a connection to the large scale poloidal field traced by the NTFs.

Our 7 new vectors in this region provide additional support for this interpretation,
but a more extensive polarimetric mapping of this cloud is necessary before a
complete interpretation is made.

\subsubsection{G0.18-0.04}

\begin{figure}
\plotone{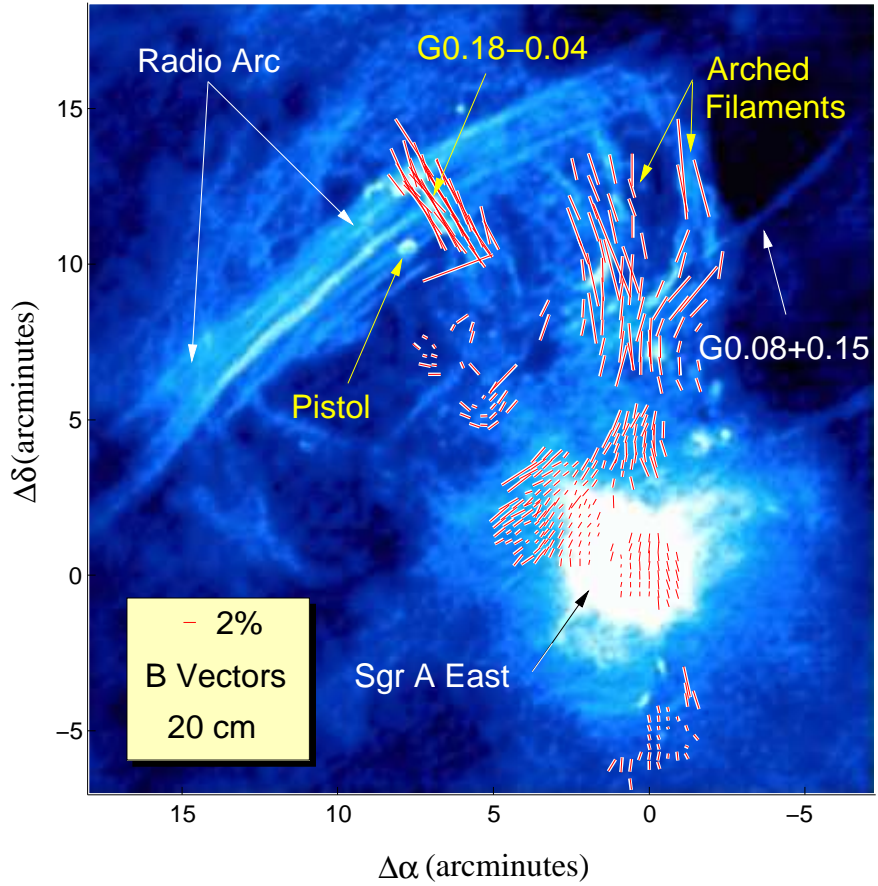}
\epsscale{1.0}
\caption{Inferred B vectors are superposed on a 20 cm map \citep{Zadeh84}
        taken with the VLA.
        Important thermal and non-thermal structures are labeled in yellow
	and white, respectively. 
        The 100 $\mu$m vectors appear to trace the Arched
        Filaments. The field in the molecular cloud associated with 
        G0.18-0.04 is perpendicular to the field associated with the   
        Radio Arc.\label{fig:20cm}}
\end{figure}

The H II region G0.18-0.04 (commonly called the ``Sickle'') can be seen in 20 cm
thermal emission in Figure~\ref{fig:20cm}. It is believed to be the ionized surface
of a molecular cloud that is interacting with the Radio Arc \citep{Serabyn91}.  Its
molecular counterpart can be seen in Figure~\ref{fig:cs}B.
It has been suggested that this interaction is responsible for producing the
relativistic electrons necessary to light up the Radio Arc via magnetic reconnection
\citep{Serabyn91,Davidson96}.

The relationship between the geometry of the cloud and its magnetic field structure
is similar to that of M-0.13-0.08. The cloud is elongated parallel to the plane of
the Galaxy and its magnetic field as inferred from far-infrared polarimetry is
parallel to the long axis. These similarities to M-0.13-0.08 suggest a similar origin
of the field geometry.

One difference between this cloud and M-0.13-0.08 is that in the case of the former
there is no observed flaring of the field.  In fact, the stark 90$^\circ$ difference
between the toroidal field observed in the molecular material and the pristine
poloidal geometry of the superposed Radio Arc suggest that there is little
connection between the toroidal field and the poloidal field in this vicinity. There
are several possibilities for why this might be the case.  First, the polarimetry
coverage may be too limited in this region. More data taken in surrounding region may
reveal a connection to a poloidal field. Second, this cloud may be more evolved than
M-0.13-0.08, and so we may be observing a magnetic field that has had enough time to
shear into a more parallel configuration than that of M-0.13-0.08. Finally, because
this molecular cloud is farther from the dynamical center of the Galaxy than is
M-0.13-0.08, the tidal forces on it are weaker, thus allowing for a more uniform
shearing along its length.

\subsubsection{M+0.11-0.08 and M+0.07-0.08}

M+0.11-0.08 and M+0.07-0.08 are two peaks of a molecular cloud complex that, like
M-0.13-0.08 and the molecular cloud associated with the Sickle, is elongated nearly
parallel to the plane with its long axis pointed toward Sgr A*. The radial velocity
of this complex has been measured to be $\sim$50 km s$^{-1}$ (see
Fig.~\ref{fig:cs}C).

The inferred magnetic field configuration for this complex displays a geometry that
is quite unique.  The field appears to tightly wrap around the southern and eastern
edges of this cloud. The implication of this curvature of the magnetic field is that
the cloud is moving toward the Sgr A* region in projection and is sweeping up
magnetic flux from the intercloud medium.  The direction of travel inferred from the
polarization data indicates that in projection, this cloud is moving toward a region
dominated by a poloidal field (see northeast side of M-0.02-0.07 in
Fig.~\ref{fig:scuba}). As the cloud moves through the less dense intercloud medium,
this poloidal flux then is wrapped around the cloud into a toroidal configuration as
seen along the eastern edge of the cloud. The result is an observed transition
between poloidal and toroidal fields.

Interior to this cloud complex, there are few measurements; however, those that exist
hint at a field that is parallel to its long axis, similar to that of M-0.13-0.08 and
the molecular cloud associated with the Sickle.

\subsubsection{The X Polarization Feature}

The 350 $\mu$m polarimetric observations of M-0.02-0.07, the CND, and CO+0.02-0.02
provide evidence for a magnetic field structure that is contiguous with that derived
from the 100 $\mu$m polarimetric observations of the Arched Filaments to form an
``X-shaped'' feature that we will refer to as the ``X Polarization Feature.'' 
The feature extends (in the coordinate system of Fig.~\ref{fig:scuba}) from 
(-1, 13) down through (5, 0). The
organization of the magnetic field vectors appears to be on scales (30 pc)
significantly larger than those of typical Galactic center molecular clouds (5-10
pc). This fact reinforces the notion that the vectors trace a single global field
that is subject to environmental forces.

It is likely that the X Polarization Feature is a line-of-sight superposition of
regions of poloidal and toroidal fields.  The orientation of the observed magnetic
field at any given location depends on the relative polarized emission from poloidal
and toroidal regions that intersect the line of sight. For example, the poloidal
field is seen to dominate in the eastern part of the M-0.02-0.07 cloud and at the
western edge of the Arched Filaments. The toroidal field dominates at the eastern
edge of the Arched Filaments and around the CND. The two fields mix in CO+0.02-0.02.

\begin{figure}
\epsscale{.8}
\plotone{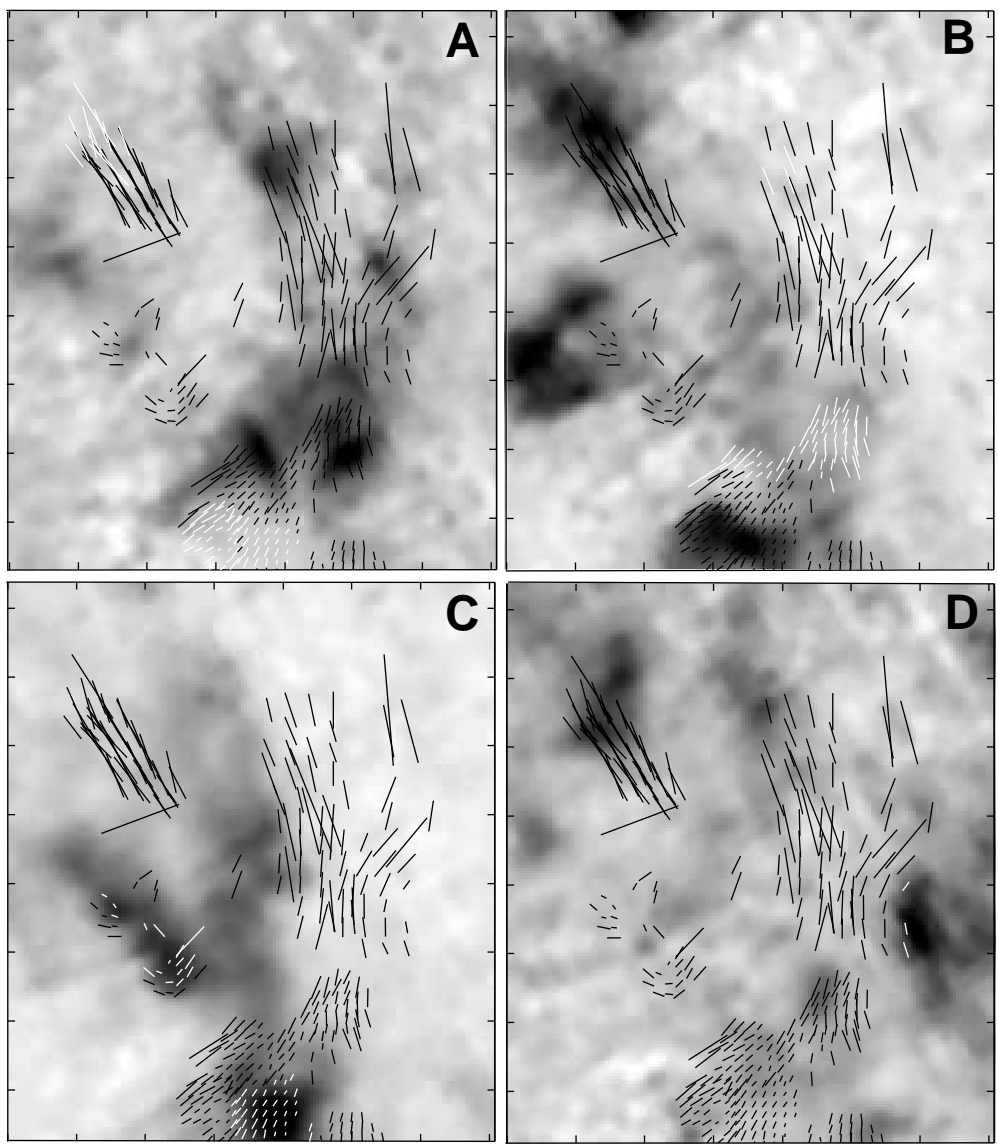}
\caption{CS maps \citep{Tsuboi97} are shown with polarization data
        superposed. Panel A shows structures at $\sim$-15 km s$^{-1}$, the
	most prominent of which is the molecular material associated with
        the arched filaments.
        Panel B shows material at $\sim$30 km s$^{-1}$.  The
        molecular material associated with the Sickle ($v\sim25$ km s$^{-1}$)
        is visible here.
        Panel C shows material at $\sim$50 km s$^{-1}$.
        M+0.11-0.08, M0.07-0.08, and M-0.02-0.07 are all prominent at this
        velocity. Note that the magnetic field vectors at the southwestern
        edge of M0.07-0.08 trace the edge of the cloud indicating that
        the cloud has a velocity in the plane of the sky that is directed
        toward M-0.02-0.07, a region of predominantly poloidal flux.
        Panel D traces $\sim$85 km s$^{-1}$ material. Note the edge of the
        cloud in the lower right of the panel seems to be traced by the  
        100 $\mu$m magnetic field vectors.
\label{fig:cs}}
\end{figure}

Another example of line-of-sight mixing of poloidal and toroidal fields occurs at the
western edge of the Arched Filaments. Here, there exist several magnetic field
vectors oriented perpendicular to the plane that are in close proximity to the
G0.08+0.15, a NTF that traces the poloidal field.  Just to the north, the magnetic
field vectors return quite abruptly to a toroidal configuration in the vicinity of
the molecular features that correspond to the Arched Filaments. It is possible that
the molecular material at $v\sim-15$ km s$^{-1}$ (See Fig.~\ref{fig:cs}A) is
displaced along the line of sight from G0.08+0.15, and the net polarization observed
consists of contributions from dust associated with the respective neighborhoods of
these two features. To the south of G0.08+0.15, the vectors again become indicative
of a field distorted by gravity as they wrap around the molecular cloud visible in
Figure~\ref{fig:cs}D.

\subsection{Polarized Flux\label{sec:polflux}}

\begin{figure}
\epsscale{1.0}
\plotone{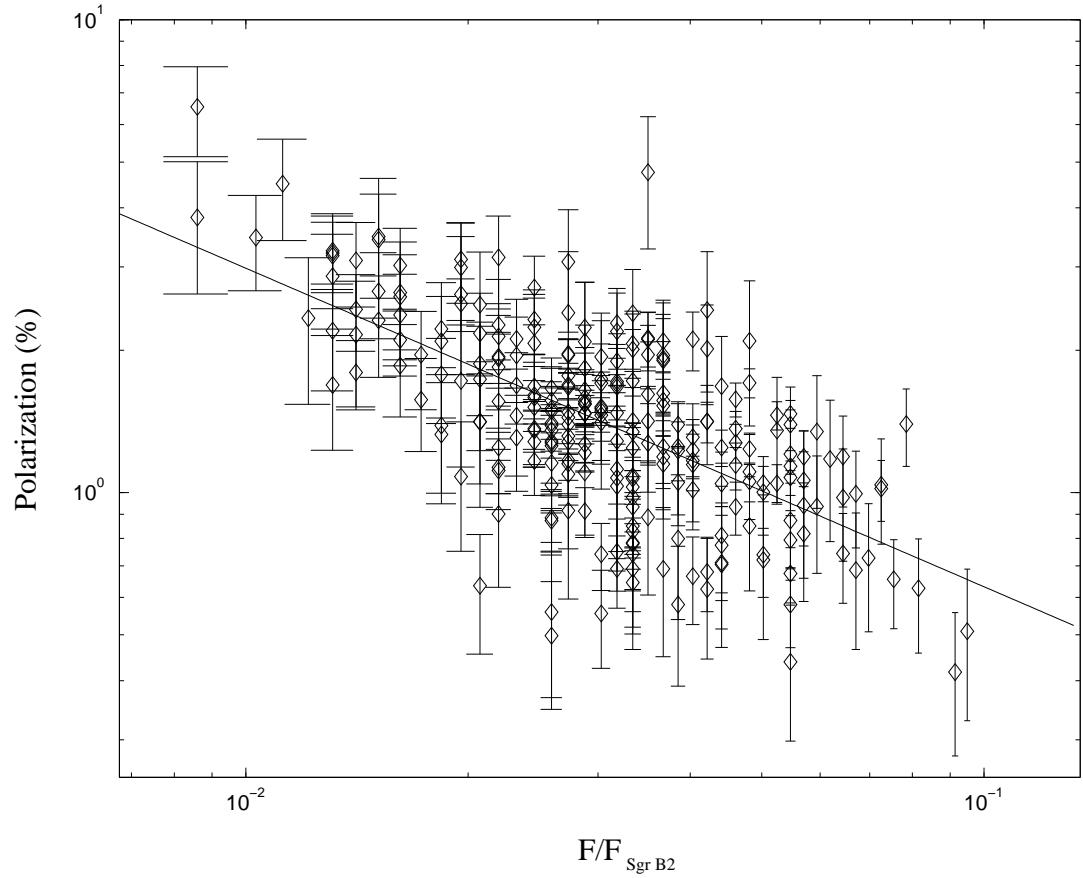}
\caption{350 \micron\, polarization is plotted against 350 \micron\,
        flux. The cutoff for these data is 3 $\sigma$ and the data
        have been corrected by $P^{\prime}=\sqrt{P^2-\sigma_P^2}$
        \citep{Serkowski74}.  The best fit slope of this curve is -0.67.
\label{fig:pvf}}
\end{figure}

Figure~\ref{fig:pvf} shows the relationship between Galactic center polarization 
measured by Hertz
versus the 350 $\mu$m flux measured by SHARC, a 350 $\mu$m photometer with a 15$\arcsec$
beam \citep{Dowell99}. A linear fit gives a
slope of -0.67. In this plot, we have included all points with a polarization
signal-to-noise (S/N) greater than 3 and have corrected the polarization by
$P^{\prime}=\sqrt{P^2-\sigma_P^2}$ in order to account for the systematic
overestimation of polarization \citep{Serkowski74} inherent in the conversion from
$q$ and $u$ to $P$ and $\phi$. Figure~\ref{fig:pvf} shows a depolarization effect
associated with an increase in 350 $\mu$m flux.

Because of a concern over a selection effect that systematically excludes low
polarizations at low fluxes, the test was repeated with a S/N cutoff of 1. For this
case, the slope decreases to -0.73, thus exonerating Figure~\ref{fig:pvf} of this
effect.  The test was repeated for the 11 individual Hertz pointing corresponding to
both this work and that of \cite{Novak00}.  A polarization versus flux comparison
using the relative fluxes obtained simultaneously with the polarizations gives an
average slope of -0.96$\pm$0.32. This number is the mean of the slopes for each of
the 11 fields. The error is the standard deviation of these slopes.

These numbers are similar to the values found by \cite{Matthews01,Matthews02} for
various Galactic molecular clouds. \cite{Matthews02} suggest three possibilities for
the depolarization effect (see also \cite{Schleuning98} and \cite{Dotson96}). First,
the lower polarization could be due to poor grain alignment or low polarization
efficiency in the cores of these clouds. Second, the depolarization could be a result
of a geometrical cancellation of the front and back of a three dimensional field that
is threaded through the optically thin cloud. Finally, it is possible that the
spatial structure of the magnetic fields in the cores of clouds is too small to be
resolved by Hertz's beam.

The evidence for depolarization in the Galactic center does not provide enough
information to differentiate among the three proposed explanations; however, it does
indicate that the depolarization effect is observable in clouds that are a factor of
$\sim$10 larger than molecular clouds in the disk of the Galaxy.

\subsection{Strength of the Magnetic Field}

Historically the lack of information concerning the degree of grain alignment and
relative strength of the line-of-sight component of the magnetic field have limited
the ability of far-infrared and submillimeter polarimetry to provide information
concerning the strengths of magnetic fields. We will show that in the Galactic center
the additional information concerning magnetic fields (i.e. the NTFs) can lead to a
model-based estimate of a characteristic global magnetic field strength estimate.

In studying the magnetic field in the central 30 pc of the Galaxy, we revisit the
model of \cite{Uchida85} as it applies to the Galactic center. Specifically, we keep
in mind that the Galactic center is clumpy and that some regions seem to contain 
a field that is poloidal while others contain a field that is toroidal.
In the spirit of this model, we adopt a picture in which gravity is assumed to
dominate the dynamics in regions of toroidal fields. Conversely, in regions of poloidal
fields, magnetic fields are assumed to be energetically dominant.

Figure~\ref{fig:phiflux} illustrates the dependence of the polarization angle on the
flux measured by SHARC \citep{Dowell99}. In this figure, the absolute deviation from
a poloidal field ($|\phi-\phi_{\mathrm{poloidal}}|$) is plotted against 350 $\mu$m
flux. This relationship shows that for regions having higher flux (and thus higher
column densities), the field is toroidal and the dynamics are gravity-dominated. For
lower intensities, poloidal fields are observed indicating that there are regions of
low density in which the dynamics are dominated by the energy density associated with
the magnetic field. Somewhere between these two extremes is a flux for which the
energy density of the poloidal field equals that of the gravitational energy density.
In order to proceed with the estimate, we will assume that this density corresponds
to a measured angle $|\phi-\phi_{\mathrm{poloidal}}|=45^\circ$. This choice is
somewhat arbitrary. In order to find the flux corresponding to this angle, we perform
various binnings as shown in Figure~\ref{fig:binned}. Linear fits to each of these
four plots give a flux of 125 Jy beam$^{-1}$. To a crude approximation, this energy
balance condition is represented by \begin{equation} \frac{1}{2}\rho v^2 =
\frac{B^2}{8\pi}. \end{equation} Here, the kinetic energy density of material
orbiting in the gravitational potential well of the Galaxy is equated to the magnetic
energy density.

The key to this problem now becomes the estimate of $\rho$ and $v$. The velocity,
$v$, can be estimated from typical cloud velocities \citep{Tsuboi99} and is expected
to be between $50-150$ km s$^{-1}$. The density can be determined as follows.

The brightness one observes from a thermal source having an optical depth $\tau_\nu$
is \begin{equation}
        I_\nu= B_\nu(T)(1-\mathrm{e}^{-\tau_\nu}),
\end{equation}
where $B_\nu(T)$ is the Planck function.

In the case of 350 $\mu$m observations, the optical depth is generally small. For
$\tau_\nu\ll1$, 
\begin{equation}
        I_\nu=\tau_\nu B_\nu(T). 
\end{equation} 
SHARC measures
$F_\nu=I_\nu\Delta\Omega$, the flux of the incoming radiation where $\Delta\Omega$ is
the solid angle subtended by a SHARC array element. From these equations, it is
possible to express the optical depth of the dust as a function of temperature and
measured flux. 
\begin{equation}
        \tau_\nu=\frac{F_\nu}{B_\nu(T)\,\Delta\Omega}
\end{equation}

Alternatively, we can express the optical depth as a function of grain properties
along the line of sight. Once again we are working in the limit $\tau_\nu\ll1$.  We
can imagine a column of dust along the line of sight that extends through the entire
depth of the Galactic center. The optical depth is proportional to the number density
of dust grains along the line of sight ($N_d$). It is also proportional to the
typical geometrical cross-section of each of the grains ($\sigma_d$); however, since
the grain sizes are generally much smaller than the wavelenth of the radiation, the
efficiencies of the grains for emitting, scattering or absorbing light are much lower
than this blackbody approximation indicates. Thus we write the optical depth as
\begin{equation}
        \tau_\nu=N_d\,Q_e\,\sigma_d, 
\end{equation}
where $Q_e$ is the emissivity of the dust grains and is generally much
less than unity for submillimeter radiation.

Once the $N_d$ is found, the total dust mass observed by a SHARC beam is
\begin{equation}
        M_d=N_d\, \rho_d\, v_d\, \Delta\Omega\, D^2. 
\end{equation} 
Here, $D$ is the
distance to the source and $\Delta\Omega D^2$ is simply the physical size of SHARC's
beam at a distance $D$. $\rho_d$ and $v_d\sim\frac{4}{3}\pi a^3$ are the density and
volume of a dust grain, respectively. If we then make the appropriate substitutions
and assume a gas-to-dust ratio, $X\gg1$, we get the following expression for the
total mass. 
\begin{equation}
        M=\frac{4\,F_\nu\, \rho_d\,a\,D^2X}{3\,Q_e\,B_\nu(T)}
\end{equation}
Putting in the appropriate numbers for 350 $\mu$m radiation yields
\begin{equation}
        \frac{M}{M_\odot}=
        \left({\frac{F}{\mathrm{Jy}}}\right)
        \left({\frac{\rho_d}{\mathrm{g\,cm^{-3}}}}\right)
        \left({\frac{a}{\mu \mathrm{m}}}\right)
        \left({\frac{D}{\mathrm{kpc}}}\right) \\
        \left({\frac{X}{Q_e}}\right)
        6.89\times10^{-7} (\mathrm{e}^{41.1/T}-1).
\end{equation}
The density can be calculated by assuming a value for the depth of the
dust layer ($L$).
\begin{equation}
        \rho=\frac{M}{\Delta\Omega^2\,D^2\,L}
\end{equation}
We use the following grain properties \citep{Dowell99} for our grain model. These are
$a=0.1\, \mu\mathrm{m}$, $Q_e=1.9\times10^{-4}$, $X=100$, and
$\rho_d=3\,\mathrm{g\,cm^{-3}}$. In addition, Pierce-Price, et al. (2000) have used
SCUBA to map the Central Molecular Zone (CMZ) at 450 and 850 $\mu$m. They have found
the dust temperature to be relatively uniform over the CMZ and adopt a value of 20 K.  
With these numbers, one can get an estimate of the magnetic field strength as a
function of velocity of the material and the thickness of the dust.
\begin{equation}
        B=3 \,\mathrm{mG}\left({\frac{L}{200\,\mathrm{pc}}}
        \right)^{-\frac{1}{2}} \left({\frac{v}{100\,
        \mathrm{km\, s^{-1}}}}\right)
\end{equation}
Based on CS measurements of the CMZ \citep{Tsuboi99}, most molecular material has v
$<$ 150 km s$^{-1}$. The CMZ has a projected diameter of 200 pc.  Assuming
cylindrical symmetry, this is approximately the scale of the material along the line
of sight.

Because of the clumpy nature of the central 30 pc, this value $L$ is most likely
overestimated, thereby making the above estimate of $B$ conservatively low.

\subsection{Reference Beam Contamination\label{sec:refbeam}}

In performing differential measurements using chopping techniques, polarized flux in
the reference beam positions is always a potential hazard. Because of the extended
nature of the dust in the Galactic center, it is of particular importance to
understand the possible effect of reference beam contamination on the data presented
here. Several attempts have been made to quantify this issue
\citep{Novak97,Schleuning97,Matthews01}. The goal of this section is to assess the
level of reference beam contamination in our data. Note that throughout this analysis
quantities pertaining to the ``reference beam'' such as $P_\mathrm{r}$ and
$I_\mathrm{r}$ refer to the average of these quantities over the multiple reference
beam positions used in the observations.

\begin{figure}
\epsscale{1.0}
\plotone{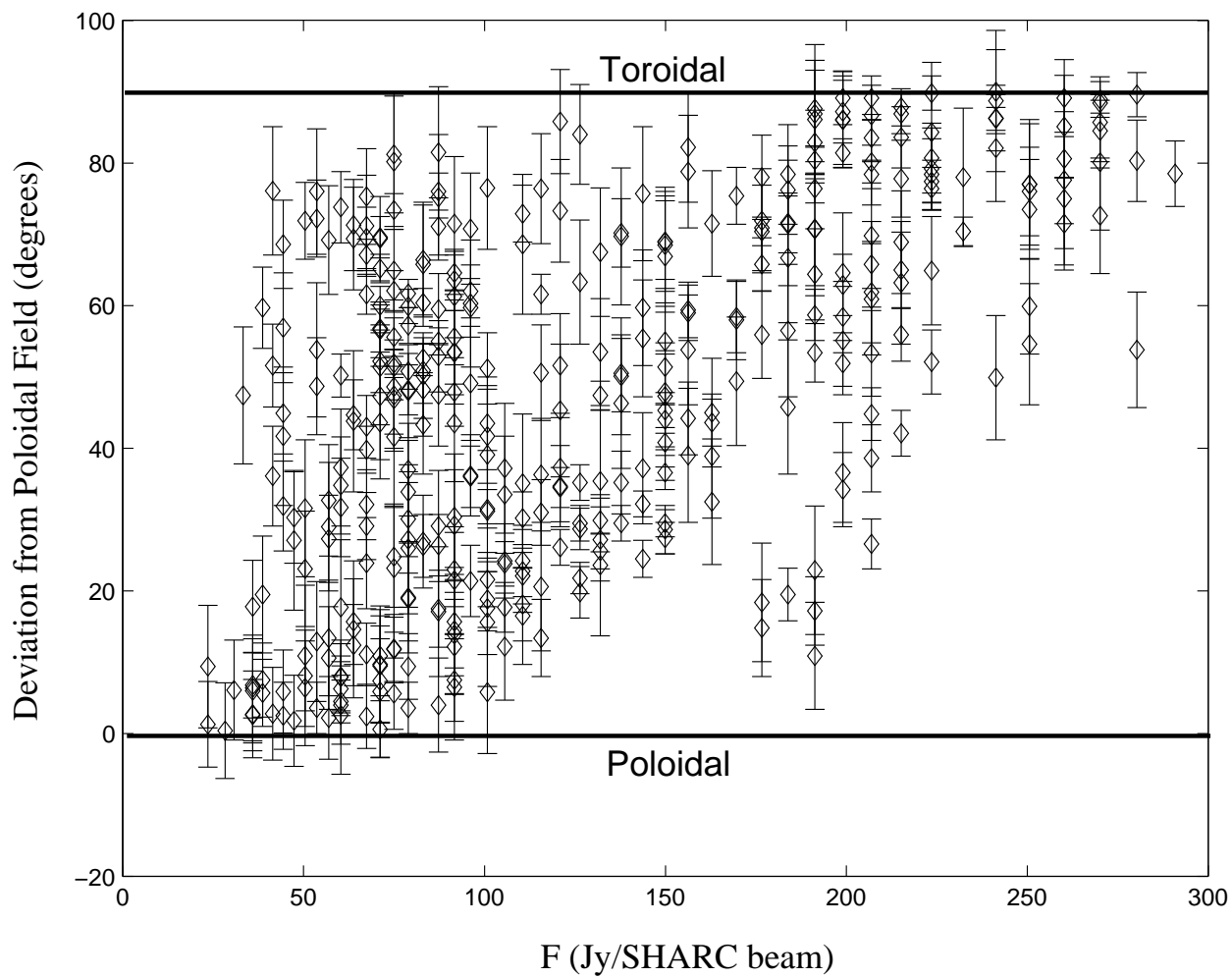}
\caption{The absolute value of the deviation of each measurement from a
        poloidal field is plotted against 350 $\mu$m flux in a
        15$^{\prime\prime}$ SHARC beam.
        Here the angles are used from all of the GC polarization measurements.
        The assumption is that the polarization angle will not change
        significantly from 60 $\mu$m to 350 $\mu$m.
\label{fig:phiflux}}
\end{figure}

Since the quantitative result of this work centers on the relationship shown in
Figure~\ref{fig:phiflux}, we are concerned primarily with the polarization angle
($\phi$), and so we wish to estimate the maximum effect of polarized flux in the
reference beam on $\phi$. The difference between the source polarization angle
($\Phi_\mathrm{s}$) and that measured ($\Phi_\mathrm{m}$) is given by \cite{Novak97}
as
\begin{equation}
        \Phi_\mathrm{s}-\Phi_\mathrm{m}=\frac{1}{2}\tan^{-1}\left[\frac{P_rw\,\sin(2\delta)}
        {\sqrt{(P_\mathrm{m}^2-P_\mathrm{r}^2w^2\,\cos(2\delta))}}\right].
\label{eq:rberr}
\end{equation}
Here, $P_\mathrm{r}$ is the polarization in the reference beam, $\delta$ is the difference
between the measured polarization angle and the polarization angle of the reference
beam ($\Phi_\mathrm{m}-\Phi_\mathrm{r}$), and $w$ is the ratio of flux in the reference beam to that
measured for a given source ($I_\mathrm{r}/I_\mathrm{m}$). Alternatively, since
$I_\mathrm{s}=I_\mathrm{m}+I_\mathrm{r}$, $w$ can be expressed as
$(I_\mathrm{s}/I_\mathrm{r}-1)^{-1}$. Here, $I_\mathrm{s}$ is the intrinsic source
flux.

\begin{figure}
\plotone{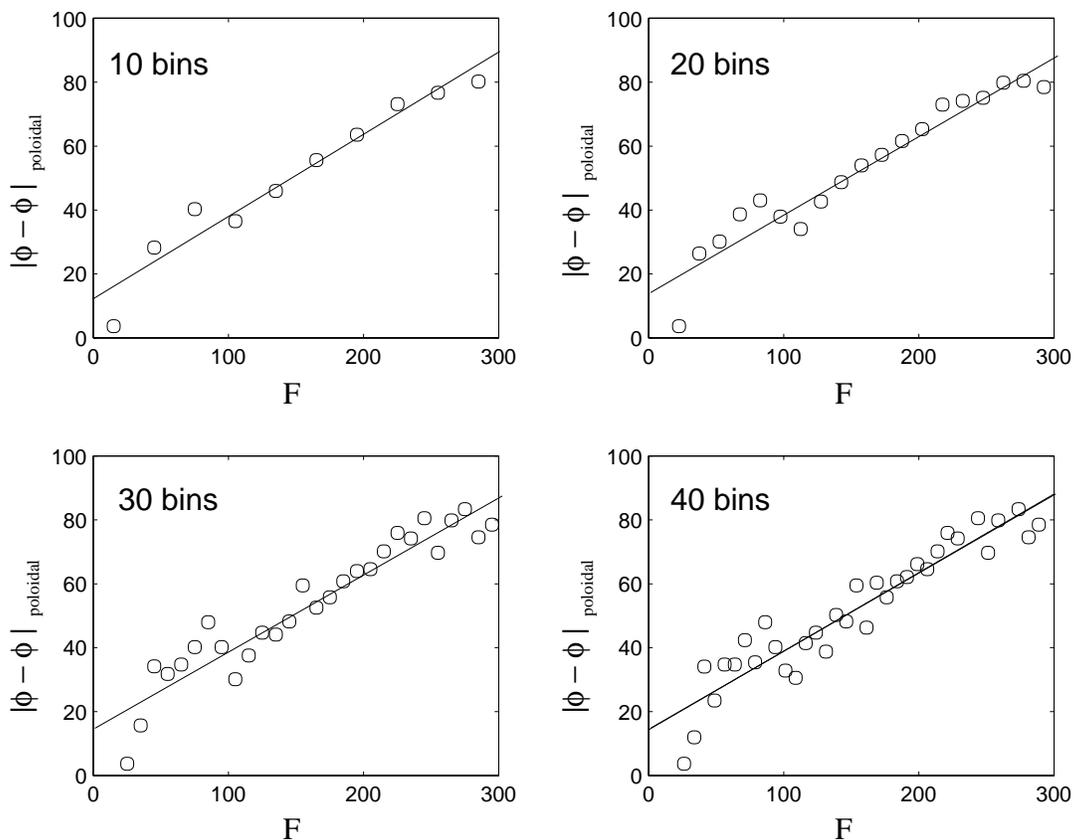}
\caption{The fluxes are grouped into bins, and the mean of polarization angle 
	in each bin is plotted against the mean flux in each bin.
        For each binning, the flux for equality
        of magnetic and gravitational forces has been calculated by fitting
        a line and then determining the flux value for a 45$^\circ$ deviation
        from a poloidal field. By taking the average of the flux values for
        each of the 4 plots, this flux is 125 Jy beam$^{-1}$. The error in the
        choice of bin size is $\sim$ 1 Jy beam$^{-1}$.
\label{fig:binned}}
\end{figure}

It is desirable to estimate the maximum effect the reference beam contamination could
have on these measurements. Specifically, we wish to understand how reference beam
contamination could affect the results shown in Figures~\ref{fig:phiflux} and
\ref{fig:binned}. To do this we need to find $w$, $P_{\mathrm{m}}$, 
$P_{\mathrm{r}}$, and $\Phi_{\mathrm{r}}$.

We find from the SCUBA survey \citep{PiercePrice00}, the average ratio of the flux in
the main beams of our six sources to that in the reference beams is
$I_\mathrm{s}/I_\mathrm{r}$=1.9. This leads to $w=1.1$.
 
The average 350 $\mu$m polarization measured by Hertz in the Galactic center is 1.9
\%. Thus, we assign $P_\mathrm{m}=0.019$.

Finally, we need to make an estimate of $P_{\mathrm{r}}$ and $\Phi_{\mathrm{r}}$, the quantities describing
the polarization properties of the reference beam flux.  To do this, we note that
\cite{Novak02} have mapped the Central Molecular Zone(CMZ) in 450 $\mu$m polarimetry.
They find the average polarization into a 6$^\prime$ beam is 1.4 \% and that the
field is uniformly toroidal. Because they use a 30$^\prime$ chop throw, their
reference beam is well off of the CMZ and the contamination should be negligible.  
Ignoring any wavelength dependence, we set $P_r$=0.014 and
$\Phi_\mathrm{r}$=-58.4$^\circ$, the polarization angle that corresponds to a
toroidal field.

\begin{figure}
\plotone{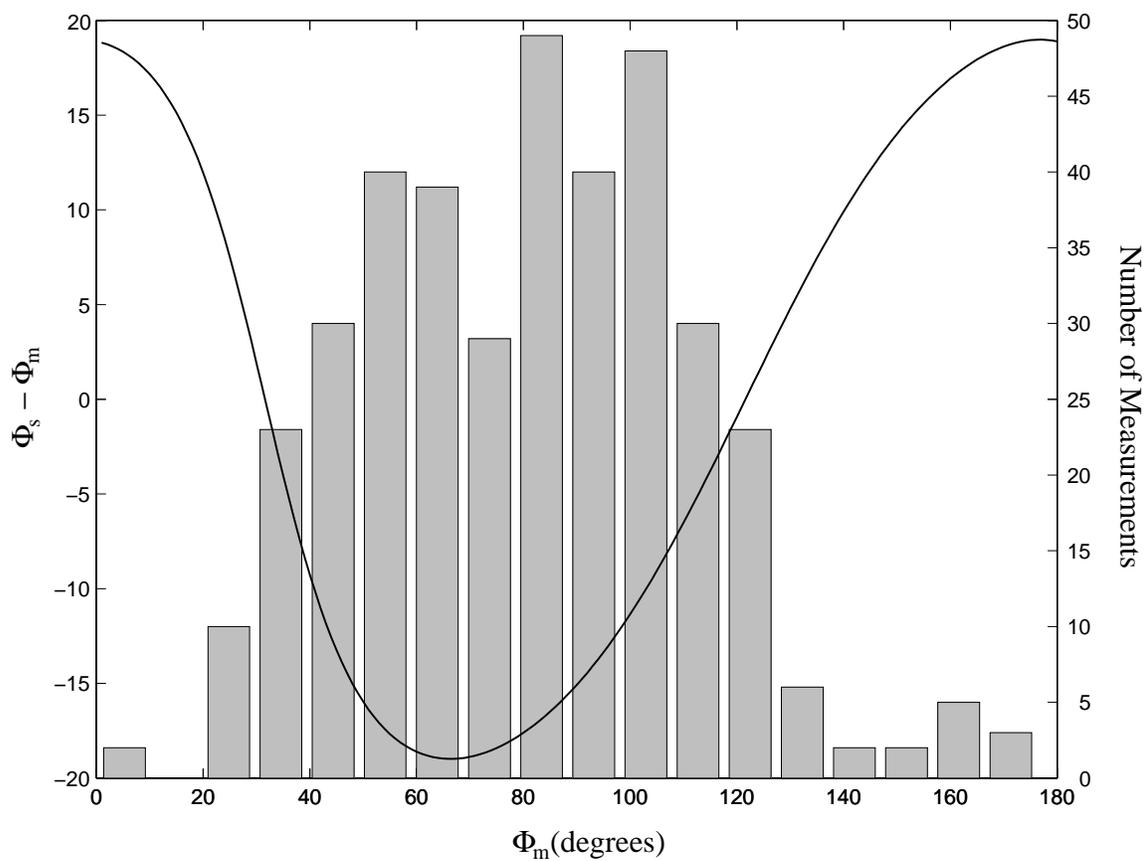}
\caption{The difference between the intrinsic source polarization angle
	and the measured polarization angle is plotted as a function of 
	measured polarization angle. This study assumes a toroidal reference
	beam polarization.  Also shown is the distribution of measured
	polarization angles used in this study.}
\label{fig:ccurve}
\end{figure}

The difference between the measured and actual polarization angle
($\Phi_\mathrm{s}-\Phi_\mathrm{m}$) under the above assumptions is plotted versus
measured polarization angle ($\Phi_\mathrm{m}$) in Figure~\ref{fig:ccurve}.  The
maximum error in ($\Phi_\mathrm{s}-\Phi_\mathrm{m}$) is $\pm$20$^\circ$. Note that
the amplitude of this error is independent of the chosen value of $\Phi_\mathrm{r}$
since the curve simply shows the error in the measured polarization angle as a
function of the difference between the measured polarization angle and the reference
beam polarization angle. (In this case, we have set the reference beam polarization
angle to an ``arbitrary'' value.) The potential error induced by reference beam
contamination is systematically about twice that of our maximum allowable statistical
error. (A signal-to-noise ratio of 3 implies $\sigma_{\phi}\sim 10^\circ$.) Though
this is potentially large, it is not large enough to cause poloidal fields to be
measured as toroidal and vice-versa.  Therefore, it is unlikely that the relationship
shown in Figures~\ref{fig:phiflux} and \ref{fig:binned} is caused by reference beam
contamination.

We can quantitatively estimate the potential effect of reference beam contamination
on the data. This histogram in Figure~\ref{fig:ccurve} shows the number of
measurements having a polarization angle in each of the 10$^\circ$ bins. As an
unfortunate coincidence, the error is not symmetric with respect to our data and thus
it is expected that there will be a bias in our field strength estimate. We have
applied this correction to the data and the resulting $\phi$ v. $F$ plots have the
same basic appearance; however the equilibrium point once the correction is applied
is 168$\pm$2 Jy (SHARC beam)$^{-1}$. This implies a 30 \% increase in our magnetic
field strength estimate.

It must be noted that because the large scale field in the Galactic center has been
found to be toroidal by \cite{Novak02}, it is possible that a significant amount of
dust is present along the line of sight to the central 30 pc that is permeated by
this toroidal field and will emit polarized radiation accordingly. In this case, the
large scale contribution to the polarized flux can be modeled by a uniform sheet of
polarized flux. In this case, chopping can be an advantage in observing magnetic
fields in the central 30 pc as it removes the contribution from a uniform field in
the foreground and background dust. This model may also explain the lack of
polarization measurements by \cite{Novak02} in the central 30 pc in regions where
Hertz sees a poloidal field.  Because of the 6\arcmin chop, Hertz may be sampling a
smaller volume along the line of sight than that of \cite{Novak02}. In the latter
case, line-of-sight superposition of the uniform sheet and the polarized emission
detected by Hertz may cancel, resulting in unpolarized radiation.

\section{Conclusions}

As described above, submillimeter and far-infrared polarimetry provides an additional
method for estimating the magnetic field strength at the Galactic center, a quantity
that has proven to be controversial. \cite{Zadeh97} have argued that the magnetic
field strength must exceed $\sim$1 mG in order to maintain the linearity of the
NTFs. On the other hand, \cite{Sofue87} have derived fields of 10-100 $\mu$G for the
Arched filaments and the Radio Arc using Faraday rotation measurements. Using equipartition
arguments, \cite{Tsuboi86}
have found a field of the same order in the plumes that make up the extension
of the Radio Arc on the eastern side of the Galactic Center Lobe. More recently,
Zeeman splitting measurements have been done in the central 30 pc both using OH masers
and H I emission \citep{Zadeh96,Plante95,Killeen92}. These studies show a line-of-sight
field of a few mG.  

Our estimate is in agreement with the notion that there is a pervasive mG field in the
Galactic center, thought obtaining a more accurate number will require a larger sample
of polarization measurements with corresponding fluxes. 
 
It has been discovered that many of the NTFs in the Galactic center are associated
with molecular clouds \citep{Serabyn91,Staguhn98}. This has led to the suggestion
that the source of the relativistic electrons in the NTFs is due to magnetic
reconnection \citep{Serabyn94}. The magnetic reconnection is believed to be
precipitated by the collision of the cloud with a magnetic flux tube either by
distorting the fields in the flux tube or by forcing these fields into contact with
those in the cloud.

\begin{figure}
\epsscale{.8}
\plotone{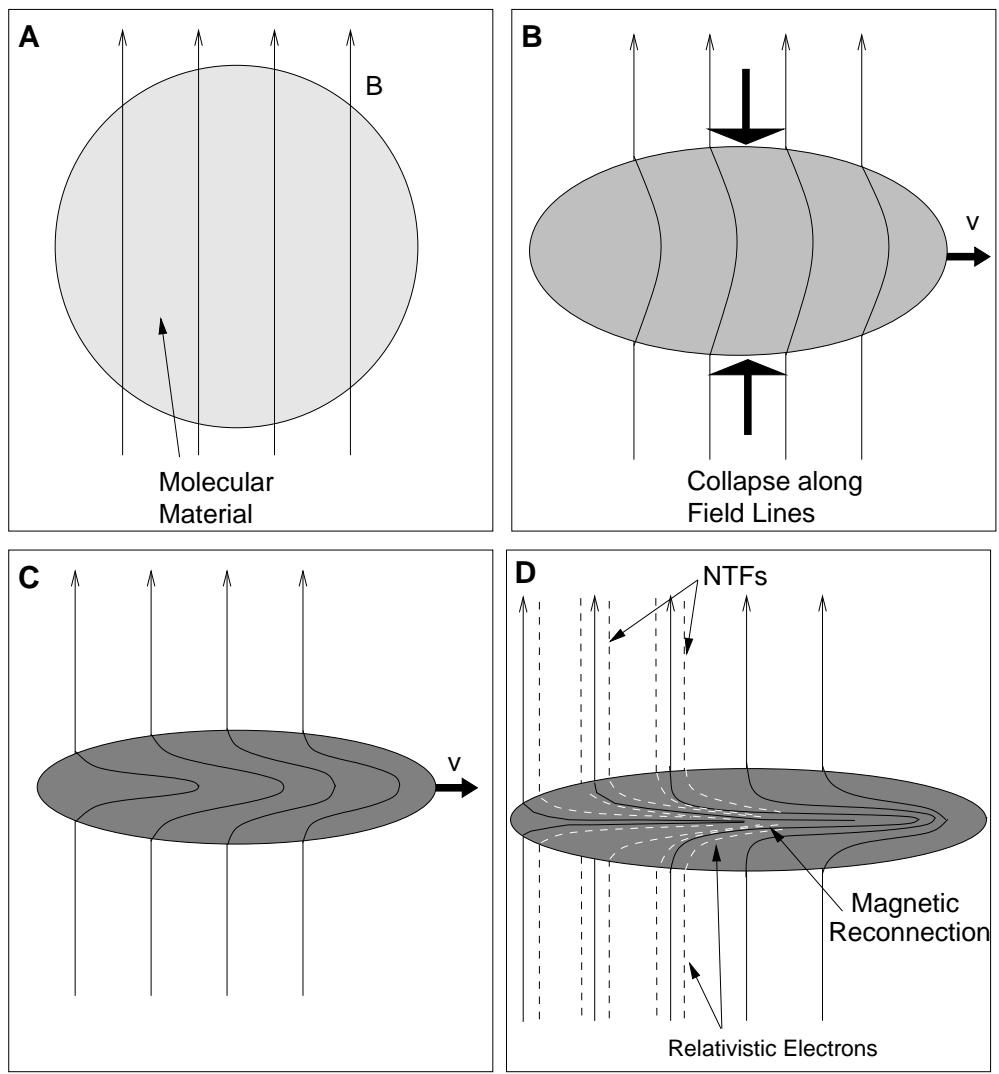}
\caption{Molecular clouds can produce relativistic electrons necessary for
        the illumination of NTFs by the following process. In regions of
        low density, the molecular material is dominated by the magnetic
        field, and we observe a poloidal field (A). MHD allows for movement of
        material along the lines of flux. In this way, the material can form
        clouds and gravity can begin to compete with the magnetic field
        energy density (B). At this stage, velocities of the molecular
        material with respect to the poloidal field can distort the field.
        This process continues (C) as the poloidal fields become sheared into
        toroidal ones in the vicinity of the cloud.  Finally,
        oppositely-oriented magnetic fields near the cloud centers will be
        forced into contact by gravity and will reconnect, thereby releasing
        energy that energizes relativistic electrons. These electrons spiral
        along the external field and produce synchrotron radiation that we
        observe as an NTF.
\label{fig:model1}}
\end{figure}

The notion that the poloidal and toroidal fields have the same origin suggests a
third option for the magnetic reconnection scenario. Figure~\ref{fig:model1}
illustrates this idea.

In the Galactic center, relatively diffuse molecular gas is supported by magnetic
pressure (Fig.~\ref{fig:model1}A); however, this material is free to collapse along
the field lines. As the gas collapses, gravitational energy becomes increasingly
important. Gravity can accelerate the cloud and the resulting differential motion
between the material in the cloud and that of the ambient medium can begin to shear
the magnetic field (Fig.~\ref{fig:model1}B). This shear continues until the field is
more toroidal than poloidal (Fig.~\ref{fig:model1}C). Note that the field external to
the clouds is still quite poloidal because magnetic energy dominates the dynamics of
this region. Finally, if the gravitational energy is large enough, and the system is
given the time to evolve, the oppositely-oriented fields in the center of the cloud
may be squeezed together to enable reconnection.  This process releases energy that
can produce the relativistic electrons required to ``light up'' the filaments.

\begin{figure}
\epsscale{.7}
\plotone{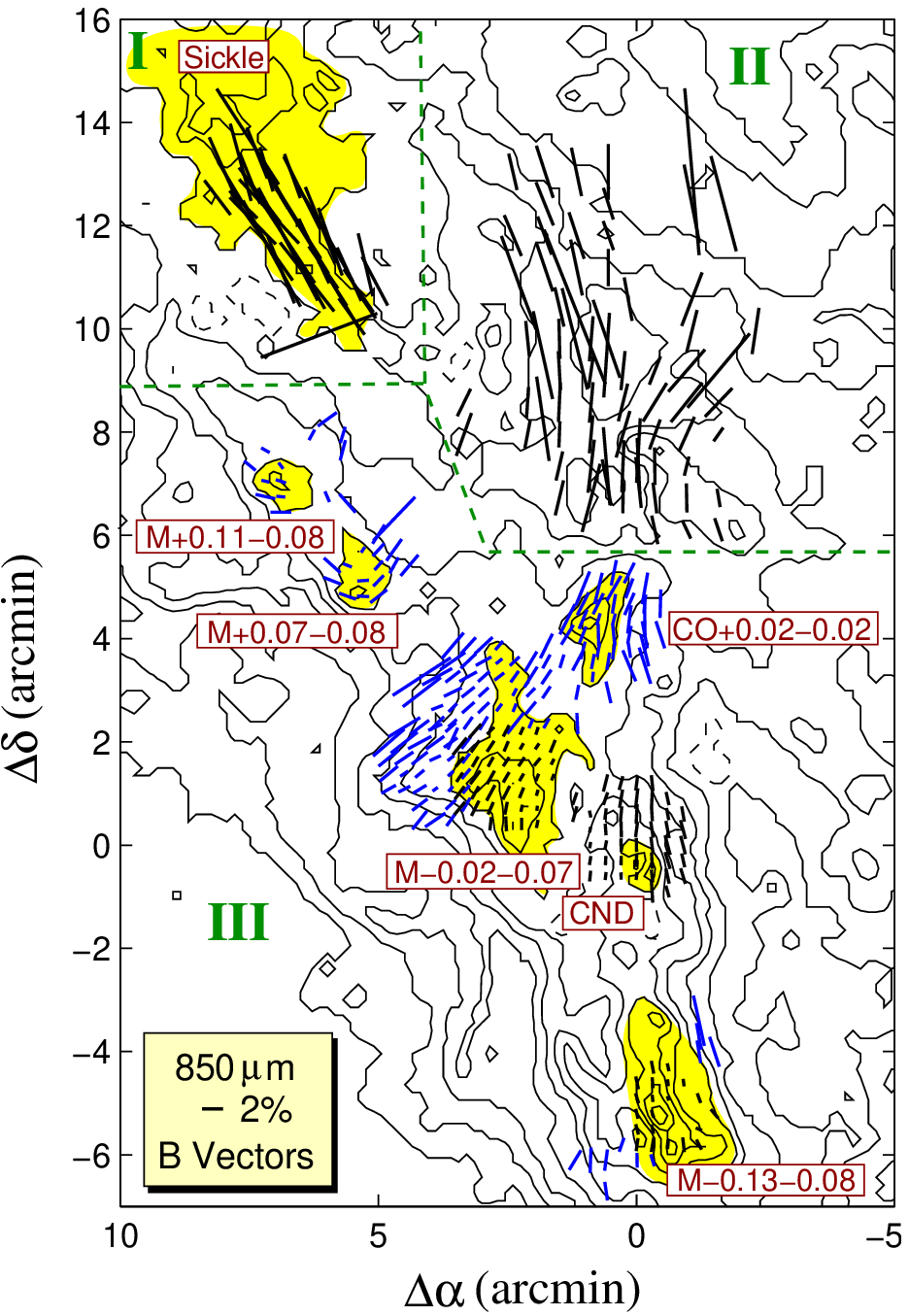}
\caption{The inferred magnetic field directions for polarization measurements
        in the Galactic center are displayed on 850 $\mu$m contours from
        SCUBA/JCMT \citep{PiercePrice00}. Region I shows 60 $\mu$m
        polarimetery of the Sickle \citep{Dotson00}. Region II shows 100
        $\mu$m polarimetry of the Arched Filaments \citep{Dotson00}.
        Region III shows new 350 $\mu$m inferred magnetic field vectors
        along with the 350 $\mu$m vectors from \cite{Novak00}.
        Important dust features are shaded and labeled. The axes scales
	are offsets in arcminutes from the position of Sgr A* ($\alpha_{2000}=
	17^{\mathrm{h}}45^{\mathrm{m}}40\fs04$, $\delta_{2000}=-29^\circ00^\prime28\farcs07$).
\label{fig:scuba}}
\end{figure}

Submillimeter and far-infrared polarimetry in concert with radio and submillimeter
photometric observations present a picture of molecular material of various sizes at
various evolutionary stages according Figure~\ref{fig:model1}. From
the polarimetry data in Figure~\ref{fig:scuba}, we can see examples of each of the
four panels in Figure~\ref{fig:model1}.  The area to the north and east of
M-0.02-0.07 is an example of the situation depicted in Figure~\ref{fig:model1}A.  
Here, the molecular material is not very dense and the magnetic field is
perpendicular to the Galactic plane. An example of Figure~\ref{fig:model1}B is the
molecular cloud complex containing M0.07-0.08 and M0.11-0.08. Here the field is seen
to be shearing around the front edge of M0.07-0.08, making a transition from poloidal
to toroidal at the southern edge of the cloud. The 20 km s$^{-1}$ cloud (M-0.13-0.08)
is thought to be shearing out its magnetic field as it falls toward Sgr A*
\citep{Novak00}. The flare seen at the southern end may indicate that it has not yet
reached the point where magnetic reconnection is occurring and hence best matches
Figure~\ref{fig:model1}C.

The interaction of the Sickle with the GCRA gives the best example of
Figure~\ref{fig:model1}D. Here, the magnetic field in the molecular cloud is observed
to be nearly perfectly aligned with the direction of both the long axis of the cloud
and the Galactic plane. Filaments, namely those of the GCRA, are observed, and appear
to diffuse into G0.18-0.04, the H II region associated with this interaction. The
clumpiness of the H II region and the structure of the filaments themselves may stem
from the irregularities of the interior of the cloud where reconnection takes place.

It is possible that there are similar occurrences in places such as the Arched
Filaments where there is also an observed transition from toroidal to poloidal fields
adjacent to an NTF (in this case, the Northern Thread). In order to test this idea 
with respect to other filaments and clouds in the Galactic center, more complete
polarimetric coverage of the central 30 parsecs is 
required.

Other scenarios for the explanation of the variety of fields in the central 30 pc
cannot yet be ruled out.  Winds produced by large explosions could be responsible for
the poloidal fields seen in this region. However, there are several reasons to favor
the association of the poloidal fields with a global field. First of all,
the spatial proximity of the poloidal fields in the dust to the NTFs lends evidence
to this association. Second, the $\sim$3 mG field strength we derive is consistent
with lower limits of the field found in NTFs. Third, to date we have not been able to
locate any sources of a wind on large enough scales that couple to the magnetic field
geometries observed. Finally, we see transitions from poloidal to toroidal fields
that correspond to dynamics of Galactic center molecular clouds, indicating that
these two fields have the same origin and that the initial configuration was
poloidal.

\acknowledgments
The authors would like to thank Farhad Yusef-Zadeh for his helpful insights. This work
was funded in part by NASA GSRP Grant number NGT 5-88.

\begin{deluxetable}{l c c c c c}
\tabletypesize{\scriptsize}
\tablecaption{M-0.13-0.08}
\tablewidth{0pt}
\tablehead{
\colhead{$\Delta\alpha$\tablenotemark{a}} &
\colhead{$\Delta\delta$\tablenotemark{a}} &
\colhead{$P(\%)$} & \colhead {$\sigma_{P}$} & \colhead {$\phi(^\circ)$} &
\colhead{$\sigma_{\phi}$}
}
\startdata
$-54 $ &$72 $ &$ 3.04 $ &$ 1.06 $ &$ 108.1 $ &$ 8.6 $  \\
$-36 $ &$72 $ &$ 2.21 $ &$ 0.71 $ &$ 99.1 $ &$ 9.0 $  \\
$-36 $ &$90 $ &$ 2.61 $ &$ 0.94 $ &$ 100.2 $ &$ 9.8 $  \\
$-36 $ &$108 $ &$ 4.98 $ &$ 1.48 $ &$ 102.4 $ &$ 7.8 $  \\
$18 $ &$-54 $ &$ 1.21 $ &$ 0.41 $ &$ 107.3 $ &$ 9.4 $  \\
$36 $ &$-54 $ &$ 1.41 $ &$ 0.40 $ &$ 91.3 $ &$ 8.1 $  \\
$54 $ &$-36 $ &$ 1.41 $ &$ 0.42 $ &$ 70.5 $ &$ 8.7 $  \\
$72 $ &$-90 $ &$ 1.83 $ &$ 0.57 $ &$ 97.4 $ &$ 8.7 $  \\
$72 $ &$-54 $ &$ 2.07 $ &$ 0.48 $ &$ 82.2 $ &$ 6.7 $  \\
$90 $ &$-54 $ &$ 2.46 $ &$ 0.57 $ &$ 87.2 $ &$ 6.6 $  \\
$108 $ &$-54 $ &$ 2.20 $ &$ 0.70 $ &$ 58.8 $ &$ 9.5 $  \\
\enddata
\tablenotetext{a}{Sky positions are relative to $17^{\mathrm{h}}45^{\mathrm{m}}37\fs30, -29\arcdeg05\arcmin39
\farcs78$ in arcseconds.}
\end{deluxetable}

\begin{deluxetable}{l c c c c c}
\tabletypesize{\scriptsize}
\tablecaption{CO+0.02-0.02}
\tablewidth{0pt}
\tablehead{
\colhead{$\Delta\alpha$\tablenotemark{a}} &
\colhead{$\Delta\delta$\tablenotemark{a}} &
\colhead{$P(\%)$} & \colhead {$\sigma_{P}$} & \colhead {$\phi(^\circ)$} &
\colhead{$\sigma_{\phi}$}
}
\startdata
$-54 $ &$-18 $ &$ 2.76 $ &$ 0.85 $ &$ 107.6 $ &$ 8.8 $  \\
$-54 $ &$18 $ &$ 2.75 $ &$ 0.74 $ &$ 88.5 $ &$ 7.7 $  \\
$-36 $ &$-54 $ &$ 3.66 $ &$ 1.14 $ &$ 107.7 $ &$ 9.0 $  \\
$-36 $ &$-36 $ &$ 3.17 $ &$ 0.60 $ &$ 103.8 $ &$ 5.4 $  \\
$-36 $ &$-18 $ &$ 2.66 $ &$ 0.59 $ &$ 100.2 $ &$ 6.2 $  \\
$-36 $ &$ 0 $ &$ 3.16 $ &$ 0.63 $ &$ 91.3 $ &$ 5.7 $  \\
$-36 $ &$18 $ &$ 1.82 $ &$ 0.59 $ &$ 85.4 $ &$ 9.4 $  \\
$-36 $ &$36 $ &$ 3.08 $ &$ 0.73 $ &$ 80.3 $ &$ 6.8 $  \\
$-18 $ &$-54 $ &$ 3.20 $ &$ 0.89 $ &$ 112.8 $ &$ 8.2 $  \\
$-18 $ &$-36 $ &$ 1.46 $ &$ 0.39 $ &$ 98.0 $ &$ 7.7 $  \\
$-18 $ &$-18 $ &$ 1.74 $ &$ 0.42 $ &$ 78.5 $ &$ 6.9 $  \\
$-18 $ &$ 0 $ &$ 2.72 $ &$ 0.55 $ &$ 83.2 $ &$ 5.8 $  \\
$-18 $ &$18 $ &$ 2.33 $ &$ 0.56 $ &$ 66.4 $ &$ 6.8 $  \\
$-18 $ &$36 $ &$ 2.02 $ &$ 0.57 $ &$ 68.9 $ &$ 8.2 $  \\
$-18 $ &$54 $ &$ 2.20 $ &$ 0.69 $ &$ 63.2 $ &$ 9.6 $  \\
$ 0 $ &$-36 $ &$ 1.74 $ &$ 0.36 $ &$ 81.8 $ &$ 6.0 $  \\
$ 0 $ &$-18 $ &$ 2.09 $ &$ 0.32 $ &$ 95.2 $ &$ 4.3 $  \\
$ 0 $ &$ 0 $ &$ 1.13 $ &$ 0.32 $ &$ 75.1 $ &$ 8.2 $  \\
$ 0 $ &$18 $ &$ 1.88 $ &$ 0.43 $ &$ 79.6 $ &$ 6.5 $  \\
$ 0 $ &$36 $ &$ 2.75 $ &$ 0.45 $ &$ 60.7 $ &$ 4.7 $  \\
$ 0 $ &$54 $ &$ 2.47 $ &$ 0.80 $ &$ 79.0 $ &$ 9.6 $  \\
$18 $ &$-36 $ &$ 1.33 $ &$ 0.34 $ &$ 79.1 $ &$ 7.2 $  \\
$18 $ &$-18 $ &$ 2.13 $ &$ 0.30 $ &$ 91.3 $ &$ 4.1 $  \\
$18 $ &$ 0 $ &$ 1.18 $ &$ 0.27 $ &$ 73.3 $ &$ 6.6 $  \\
$18 $ &$18 $ &$ 2.27 $ &$ 0.39 $ &$ 63.7 $ &$ 4.9 $  \\
$18 $ &$36 $ &$ 3.08 $ &$ 0.60 $ &$ 76.5 $ &$ 5.7 $  \\
$36 $ &$-36 $ &$ 1.71 $ &$ 0.59 $ &$ 75.4 $ &$ 9.9 $  \\
$36 $ &$-18 $ &$ 1.69 $ &$ 0.38 $ &$ 68.6 $ &$ 6.5 $  \\
$36 $ &$ 0 $ &$ 1.98 $ &$ 0.30 $ &$ 67.6 $ &$ 4.3 $  \\
$36 $ &$18 $ &$ 2.43 $ &$ 0.52 $ &$ 63.5 $ &$ 6.3 $  \\
$36 $ &$36 $ &$ 3.53 $ &$ 0.85 $ &$ 67.7 $ &$ 7.0 $  \\
$54 $ &$-18 $ &$ 2.23 $ &$ 0.75 $ &$ 58.7 $ &$ 9.8 $  \\
$54 $ &$ 0 $ &$ 3.22 $ &$ 0.70 $ &$ 63.3 $ &$ 6.9 $  \\
\enddata
\tablenotetext{a}{Sky positions are relative to 
$17^{\mathrm{h}}45^{\mathrm{m}}42\fs10, 
-28\arcdeg56\arcmin5\farcs1$ in arcseconds.}
\end{deluxetable}

\begin{deluxetable}{l c c c c c}
\tabletypesize{\scriptsize}
\tablecaption{NW of M-0.02-0.07}
\tablewidth{0pt}
\tablehead{
\colhead{$\Delta\alpha$\tablenotemark{a}} &
\colhead{$\Delta\delta$\tablenotemark{a}} &
\colhead{$P(\%)$} & \colhead {$\sigma_{P}$} & \colhead {$\phi(^\circ)$} &
\colhead{$\sigma_{\phi}$}
}
\startdata
$-54 $ &$ 0 $ &$ 2.13 $ &$ 0.51 $ &$ 102.8 $ &$ 7.1 $  \\
$-54 $ &$36 $ &$ 1.68 $ &$ 0.47 $ &$ 80.7 $ &$ 8.0 $  \\
$-36 $ &$18 $ &$ 0.97 $ &$ 0.32 $ &$ 83.0 $ &$ 9.4 $  \\
$-36 $ &$36 $ &$ 1.45 $ &$ 0.40 $ &$ 79.0 $ &$ 7.8 $  \\
$-18 $ &$-36 $ &$ 2.05 $ &$ 0.41 $ &$ 92.9 $ &$ 5.8 $  \\
$-18 $ &$ 0 $ &$ 1.46 $ &$ 0.46 $ &$ 78.9 $ &$ 8.9 $  \\
$-18 $ &$36 $ &$ 2.01 $ &$ 0.46 $ &$ 61.9 $ &$ 6.4 $  \\
$ 0 $ &$36 $ &$ 1.33 $ &$ 0.46 $ &$ 73.3 $ &$ 9.7 $  \\
$18 $ &$-36 $ &$ 1.07 $ &$ 0.30 $ &$ 60.8 $ &$ 7.8 $  \\
$18 $ &$ 0 $ &$ 1.36 $ &$ 0.34 $ &$ 56.4 $ &$ 7.1 $  \\
$18 $ &$18 $ &$ 1.63 $ &$ 0.37 $ &$ 55.5 $ &$ 6.4 $  \\
$18 $ &$36 $ &$ 1.49 $ &$ 0.48 $ &$ 60.6 $ &$ 9.0 $  \\
$18 $ &$54 $ &$ 2.60 $ &$ 0.73 $ &$ 64.3 $ &$ 7.8 $  \\
$36 $ &$-36 $ &$ 0.98 $ &$ 0.34 $ &$ 66.7 $ &$ 9.7 $  \\
$36 $ &$-18 $ &$ 1.50 $ &$ 0.31 $ &$ 50.5 $ &$ 5.9 $  \\
$36 $ &$ 0 $ &$ 1.12 $ &$ 0.33 $ &$ 45.5 $ &$ 8.4 $  \\
$36 $ &$18 $ &$ 2.31 $ &$ 0.36 $ &$ 58.0 $ &$ 4.3 $  \\
$36 $ &$36 $ &$ 2.22 $ &$ 0.47 $ &$ 58.9 $ &$ 6.1 $  \\
$54 $ &$-36 $ &$ 2.56 $ &$ 0.80 $ &$ 52.2 $ &$ 9.0 $  \\
\enddata
\tablenotetext{a}{Sky positions are relative to 
$17^{\mathrm{h}}45^{\mathrm{m}}46\fs7, 
-28\arcdeg57\arcmin30\farcs0$ in arcseconds.}
\end{deluxetable}

\begin{deluxetable}{l c c c c c}
\tabletypesize{\scriptsize}
\tablecaption{NE of M-0.02-0.07}
\tablewidth{0pt}
\tablehead{
\colhead{$\Delta\alpha$\tablenotemark{a}} &
\colhead{$\Delta\delta$\tablenotemark{a}} &
\colhead{$P(\%)$} & \colhead {$\sigma_{P}$} & \colhead {$\phi(^\circ)$} &
\colhead{$\sigma_{\phi}$}
}
\startdata
$-54 $ &$-18 $ &$ 0.84 $ &$ 0.26 $ &$ 68.8 $ &$ 9.1 $  \\
$-54 $ &$ 0 $ &$ 0.78 $ &$ 0.24 $ &$ 62.0 $ &$ 8.8 $  \\
$-54 $ &$18 $ &$ 1.33 $ &$ 0.23 $ &$ 55.8 $ &$ 5.0 $  \\
$-54 $ &$36 $ &$ 1.27 $ &$ 0.33 $ &$ 43.8 $ &$ 7.5 $  \\
$-36 $ &$-18 $ &$ 0.88 $ &$ 0.23 $ &$ 53.1 $ &$ 7.5 $  \\
$-36 $ &$18 $ &$ 1.17 $ &$ 0.21 $ &$ 37.2 $ &$ 5.0 $  \\
$-36 $ &$36 $ &$ 0.81 $ &$ 0.22 $ &$ 38.1 $ &$ 7.4 $  \\
$-36 $ &$54 $ &$ 1.13 $ &$ 0.35 $ &$ 48.7 $ &$ 9.1 $  \\
$-18 $ &$-36 $ &$ 1.52 $ &$ 0.22 $ &$ 57.9 $ &$ 4.4 $  \\
$-18 $ &$-18 $ &$ 1.46 $ &$ 0.23 $ &$ 43.4 $ &$ 4.5 $  \\
$-18 $ &$ 0 $ &$ 1.28 $ &$ 0.25 $ &$ 42.5 $ &$ 5.6 $  \\
$-18 $ &$18 $ &$ 1.46 $ &$ 0.37 $ &$ 42.2 $ &$ 7.2 $  \\
$-18 $ &$36 $ &$ 1.70 $ &$ 0.28 $ &$ 29.2 $ &$ 4.5 $  \\
$-18 $ &$54 $ &$ 2.35 $ &$ 0.36 $ &$ 42.7 $ &$ 4.4 $  \\
$ 0 $ &$-36 $ &$ 1.44 $ &$ 0.25 $ &$ 54.6 $ &$ 5.2 $  \\
$ 0 $ &$-18 $ &$ 1.58 $ &$ 0.26 $ &$ 41.0 $ &$ 4.6 $  \\
$ 0 $ &$ 0 $ &$ 1.58 $ &$ 0.26 $ &$ 39.5 $ &$ 4.7 $  \\
$ 0 $ &$18 $ &$ 1.92 $ &$ 0.42 $ &$ 29.4 $ &$ 5.8 $  \\
$ 0 $ &$36 $ &$ 1.99 $ &$ 0.40 $ &$ 44.0 $ &$ 5.7 $  \\
$ 0 $ &$54 $ &$ 2.16 $ &$ 0.45 $ &$ 47.3 $ &$ 6.0 $  \\
$18 $ &$-54 $ &$ 2.30 $ &$ 0.55 $ &$ 65.5 $ &$ 7.2 $  \\
$18 $ &$-36 $ &$ 1.08 $ &$ 0.30 $ &$ 41.1 $ &$ 7.9 $  \\
$18 $ &$-18 $ &$ 1.13 $ &$ 0.33 $ &$ 44.5 $ &$ 8.4 $  \\
$18 $ &$ 0 $ &$ 1.90 $ &$ 0.41 $ &$ 25.7 $ &$ 5.8 $  \\
$18 $ &$18 $ &$ 1.75 $ &$ 0.46 $ &$ 38.0 $ &$ 7.4 $  \\
$18 $ &$36 $ &$ 2.94 $ &$ 0.65 $ &$ 34.3 $ &$ 6.1 $  \\
$18 $ &$54 $ &$ 3.25 $ &$ 0.72 $ &$ 49.4 $ &$ 6.5 $  \\
$36 $ &$-36 $ &$ 1.81 $ &$ 0.51 $ &$ 45.0 $ &$ 8.1 $  \\
$36 $ &$-18 $ &$ 1.20 $ &$ 0.44 $ &$ 20.7 $ &$ 9.9 $  \\
$36 $ &$18 $ &$ 3.30 $ &$ 0.60 $ &$ 29.0 $ &$ 5.0 $  \\
$54 $ &$ 0 $ &$ 4.63 $ &$ 1.09 $ &$ 37.7 $ &$ 7.0 $  \\
$54 $ &$18 $ &$ 6.69 $ &$ 1.41 $ &$ 32.9 $ &$ 6.0 $  \\
\enddata
\tablenotetext{a}{Sky positions are relative to 
$17^{\mathrm{h}}45^{\mathrm{m}}55\fs0, 
-28\arcdeg57\arcmin29\farcs7$ in arcseconds.}
\end{deluxetable}

\begin{deluxetable}{l c c c c c}
\tabletypesize{\scriptsize}
\tablecaption{E of M-0.02-0.07}
\tablewidth{0pt}
\tablehead{
\colhead{$\Delta\alpha$\tablenotemark{a}} &
\colhead{$\Delta\delta$\tablenotemark{a}} &
\colhead{$P(\%)$} & \colhead {$\sigma_{P}$} & \colhead {$\phi(^\circ)$} &
\colhead{$\sigma_{\phi}$}
}
\startdata
$-54 $ &$-36 $ &$ 1.49 $ &$ 0.31 $ &$ 58.5 $ &$ 6.5 $  \\
$-54 $ &$-18 $ &$ 1.03 $ &$ 0.18 $ &$ 49.7 $ &$ 5.1 $  \\
$-54 $ &$ 0 $ &$ 1.26 $ &$ 0.16 $ &$ 49.3 $ &$ 3.5 $  \\
$-54 $ &$18 $ &$ 1.18 $ &$ 0.16 $ &$ 53.7 $ &$ 3.9 $  \\
$-54 $ &$36 $ &$ 1.44 $ &$ 0.26 $ &$ 53.0 $ &$ 5.0 $  \\
$-36 $ &$-54 $ &$ 1.38 $ &$ 0.38 $ &$ 54.7 $ &$ 8.1 $  \\
$-36 $ &$-36 $ &$ 1.30 $ &$ 0.22 $ &$ 43.5 $ &$ 4.8 $  \\
$-36 $ &$-18 $ &$ 0.67 $ &$ 0.18 $ &$ 47.3 $ &$ 7.6 $  \\
$-36 $ &$ 0 $ &$ 0.71 $ &$ 0.17 $ &$ 43.8 $ &$ 6.6 $  \\
$-36 $ &$18 $ &$ 0.80 $ &$ 0.16 $ &$ 39.1 $ &$ 5.8 $  \\
$-36 $ &$36 $ &$ 1.60 $ &$ 0.21 $ &$ 35.2 $ &$ 3.6 $  \\
$-36 $ &$54 $ &$ 1.34 $ &$ 0.38 $ &$ 41.3 $ &$ 8.2 $  \\
$-18 $ &$-54 $ &$ 2.38 $ &$ 0.56 $ &$ 34.4 $ &$ 6.5 $  \\
$-18 $ &$-18 $ &$ 0.59 $ &$ 0.19 $ &$ 37.5 $ &$ 9.2 $  \\
$-18 $ &$18 $ &$ 1.18 $ &$ 0.24 $ &$ 24.2 $ &$ 5.9 $  \\
$-18 $ &$36 $ &$ 2.15 $ &$ 0.23 $ &$ 36.1 $ &$ 3.0 $  \\
$-18 $ &$54 $ &$ 1.44 $ &$ 0.39 $ &$ 25.2 $ &$ 8.1 $  \\
$ 0 $ &$-54 $ &$ 2.25 $ &$ 0.64 $ &$ 51.1 $ &$ 8.2 $  \\
$ 0 $ &$-36 $ &$ 1.82 $ &$ 0.30 $ &$ 37.3 $ &$ 4.7 $  \\
$ 0 $ &$-18 $ &$ 1.52 $ &$ 0.21 $ &$ 31.0 $ &$ 4.0 $  \\
$ 0 $ &$ 0 $ &$ 1.15 $ &$ 0.21 $ &$ 39.6 $ &$ 5.1 $  \\
$ 0 $ &$18 $ &$ 1.94 $ &$ 0.26 $ &$ 37.9 $ &$ 3.8 $  \\
$ 0 $ &$36 $ &$ 2.53 $ &$ 0.32 $ &$ 28.0 $ &$ 3.6 $  \\
$ 0 $ &$54 $ &$ 2.27 $ &$ 0.56 $ &$ 37.6 $ &$ 7.3 $  \\
$18 $ &$-18 $ &$ 1.27 $ &$ 0.25 $ &$ 27.6 $ &$ 5.5 $  \\
$18 $ &$ 0 $ &$ 0.94 $ &$ 0.27 $ &$ 34.2 $ &$ 8.3 $  \\
$18 $ &$18 $ &$ 2.13 $ &$ 0.34 $ &$ 34.1 $ &$ 4.7 $  \\
$18 $ &$36 $ &$ 3.25 $ &$ 0.52 $ &$ 38.4 $ &$ 4.6 $  \\
$18 $ &$54 $ &$ 3.55 $ &$ 0.79 $ &$ 31.2 $ &$ 6.7 $  \\
$36 $ &$-18 $ &$ 1.61 $ &$ 0.35 $ &$ 29.8 $ &$ 6.4 $  \\
$36 $ &$ 0 $ &$ 2.25 $ &$ 0.38 $ &$ 39.7 $ &$ 4.9 $  \\
$36 $ &$18 $ &$ 2.48 $ &$ 0.45 $ &$ 39.1 $ &$ 5.2 $  \\
$36 $ &$36 $ &$ 4.00 $ &$ 1.19 $ &$ 41.0 $ &$ 8.6 $  \\
\enddata
\tablenotetext{a}{Sky positions are relative to 
$17^{\mathrm{h}}45^{\mathrm{m}}59\fs3, 
-28\arcdeg59\arcmin4\farcs5$ in arcseconds.}
\end{deluxetable}

\begin{deluxetable}{l c c c c c}
\tabletypesize{\scriptsize}
\tablecaption{M+0.07-0.08}
\tablewidth{0pt}
\tablehead{
\colhead{$\Delta\alpha$\tablenotemark{a}} &
\colhead{$\Delta\delta$\tablenotemark{a}} &
\colhead{$P(\%)$} & \colhead {$\sigma_{P}$} & \colhead {$\phi(^\circ)$} &
\colhead{$\sigma_{\phi}$}
}
\startdata
$-54 $ &$-18 $ &$ 2.64 $ &$ 0.88 $ &$ 49.3 $ &$ 9.5 $  \\
$-36 $ &$-18 $ &$ 1.76 $ &$ 0.47 $ &$ 60.7 $ &$ 7.8 $  \\
$-36 $ &$ 0 $ &$ 2.54 $ &$ 0.83 $ &$ 54.8 $ &$ 9.0 $  \\
$-36 $ &$36 $ &$ 5.82 $ &$ 1.97 $ &$ 46.2 $ &$ 9.6 $  \\
$-18 $ &$-54 $ &$ 2.04 $ &$ 0.60 $ &$ 37.4 $ &$ 8.6 $  \\
$-18 $ &$-36 $ &$ 1.44 $ &$ 0.27 $ &$ 45.0 $ &$ 5.4 $  \\
$-18 $ &$-18 $ &$ 1.18 $ &$ 0.28 $ &$ 48.1 $ &$ 6.8 $  \\
$-18 $ &$ 0 $ &$ 1.65 $ &$ 0.52 $ &$ 47.2 $ &$ 9.0 $  \\
$-18 $ &$18 $ &$ 2.16 $ &$ 0.76 $ &$ 57.6 $ &$ 9.2 $  \\
$ 0 $ &$-54 $ &$ 1.17 $ &$ 0.40 $ &$ 176.2 $ &$ 9.9 $  \\
$ 0 $ &$-36 $ &$ 1.13 $ &$ 0.23 $ &$ 4.4 $ &$ 5.8 $  \\
$ 0 $ &$ 0 $ &$ 0.62 $ &$ 0.22 $ &$ 55.2 $ &$ 9.9 $  \\
$18 $ &$-54 $ &$ 1.74 $ &$ 0.46 $ &$ 160.0 $ &$ 7.3 $  \\
$18 $ &$-18 $ &$ 0.88 $ &$ 0.23 $ &$ 158.1 $ &$ 7.5 $  \\
$18 $ &$54 $ &$ 2.39 $ &$ 0.80 $ &$ 131.4 $ &$ 9.5 $  \\
$36 $ &$-36 $ &$ 1.79 $ &$ 0.49 $ &$ 158.1 $ &$ 7.3 $  \\
$36 $ &$-18 $ &$ 2.13 $ &$ 0.30 $ &$ 138.7 $ &$ 4.0 $  \\
\enddata
\tablenotetext{a}{Sky positions are relative to 
$17^{\mathrm{h}}46^{\mathrm{m}}4\fs3, 
-28\arcdeg54\arcmin44\farcs3$ in arcseconds.}
\end{deluxetable}

\begin{deluxetable}{l c c c c c}
\tabletypesize{\scriptsize}
\tablecaption{M+0.11-0.08}
\tablewidth{0pt}
\tablehead{
\colhead{$\Delta\alpha$\tablenotemark{a}} &
\colhead{$\Delta\delta$\tablenotemark{a}} &
\colhead{$P(\%)$} & \colhead {$\sigma_{P}$} & \colhead {$\phi(^\circ)$} &
\colhead{$\sigma_{\phi}$}
}
\startdata
$-54 $ &$18 $ &$ 1.99 $ &$ 0.46 $ &$ 74.9 $ &$ 6.9 $  \\
$-54 $ &$36 $ &$ 1.62 $ &$ 0.55 $ &$ 73.2 $ &$ 9.9 $  \\
$-36 $ &$-36 $ &$ 1.09 $ &$ 0.35 $ &$ 113.1 $ &$ 9.2 $  \\
$-36 $ &$54 $ &$ 2.25 $ &$ 0.51 $ &$ 35.6 $ &$ 6.6 $  \\
$-18 $ &$36 $ &$ 1.14 $ &$ 0.29 $ &$ 61.7 $ &$ 7.6 $  \\
$18 $ &$-54 $ &$ 2.01 $ &$ 0.60 $ &$ 0.4 $ &$ 8.5 $  \\
$18 $ &$-36 $ &$ 0.84 $ &$ 0.22 $ &$ 174.3 $ &$ 7.6 $  \\
$18 $ &$-18 $ &$ 1.02 $ &$ 0.19 $ &$ 161.1 $ &$ 5.4 $  \\
$18 $ &$ 0 $ &$ 0.73 $ &$ 0.19 $ &$ 125.8 $ &$ 7.3 $  \\
$36 $ &$-36 $ &$ 1.95 $ &$ 0.43 $ &$ 168.1 $ &$ 6.5 $  \\
$36 $ &$-18 $ &$ 0.75 $ &$ 0.24 $ &$ 166.3 $ &$ 9.3 $  \\
$36 $ &$18 $ &$ 0.93 $ &$ 0.28 $ &$ 151.3 $ &$ 8.9 $  \\
$54 $ &$ 0 $ &$ 1.42 $ &$ 0.46 $ &$ 140.1 $ &$ 9.4 $  \\
\enddata
\tablenotetext{a}{Sky positions are relative to 
$17^{\mathrm{h}}46^{\mathrm{m}}10\fs2, 
-28\arcdeg53\arcmin6\farcs3$ in arcseconds.}
\end{deluxetable}

\end{document}